\documentclass[10pt,aps,physrev,reprint,floatfix,superscriptaddress,longbibliography]{revtex4-2}
\usepackage{graphicx}% Include figure files
\usepackage{dcolumn}% Align table columns on decimal point
\usepackage{bm}% bold math
\usepackage{verbatim}   % for math
\usepackage[usenames]{color}  % for color words
\usepackage{stmaryrd}
\usepackage{float}
\usepackage{units}
\usepackage{textcomp}
\usepackage{amsmath}
\usepackage{commath}
\usepackage{amssymb}
\usepackage{esint}
\usepackage{ae,aecompl}
\usepackage[colorlinks=true,linkcolor=blue,citecolor=blue]{hyperref}
\usepackage{subfigure}
\usepackage[utf8]{inputenc}
\usepackage{array}
\begin{document}

\title{Hofstadter-like spectrum and Magnetization of Artificial Graphene \\ constructed with cylindrical and elliptical quantum dots}

\author{Maryam Mansoury}
\affiliation{Department of Physics, Urmia University of Technology, Urmia, Iran}
%\email{mansourymaryam.uut@gmail.com}

\author{Vram Mughnetsyan}
\affiliation{Department of Solid State Physics, Yerevan State University, Alex Manoogian 1, 0025 Yerevan, Armenia}
\email{vram@ysu.am}

\author{Aram Manaselyan}
\affiliation{Department of Solid State Physics, Yerevan State University, Alex Manoogian 1, 0025 Yerevan, Armenia}
\email{vram@ysu.am}

\author{Albert Kirakosyan}
\affiliation{Department of Solid State Physics, Yerevan State University, Alex Manoogian 1, 0025 Yerevan, Armenia}
%\email{kirakosyan@ysu.am}

\author{Vidar Gudmundsson}
\affiliation{Science Institute, University of Iceland, Dunhaga 3, IS-107 Reykjavik, Iceland}
%\email{vidar@hi.is}

\author{Vigen Aziz-Aghchegala}
\affiliation{Department of Physics, Urmia University of Technology, Urmia, Iran}
%\email{...}

\begin{abstract}
In this paper a comparative study of the electronic and magnetic properties of quasi-two-dimensional electrons in an artificial graphene-like superlattice composed of circular and elliptical
quantum dots is presented. A complete orthonormal set of basis wave functions, which has pre\-viously been constructed in the frame of the Coulomb gauge for the vector potential has been implemented
for calculation of the energy dispersions, the Hofstadter spectra, the density of states and the orbital magnetization of the considered systems, taking into account both the translational symmetry of the
superlattice and the wave function phase-shifts due to the presence of a transverse external magnetic field.
Our calculations indicate a topological change in the miniband structure due to the ellipticity
of the quantum dots, and non-trivial modifications of the electron energy dispersion surfaces in
reciprocal space with the change of the number of magnetic flux quanta through the unit cell of
the superlattice. The ellipticity of the QDs leads to an opening of a gap and considerable modifications
of the Hofstadter spectrum. The orbital magnetization is shown to reveal significant oscillations
with the change of the magnetic flux. The deviation from the circular geometry of quantum dots
has a qualitative impact on the dependencies of the magnetization on both the magnetic flux and
the temperature.  
\end{abstract}

\maketitle
\section{INTRODUCTION}
The unique properties of graphene, which are a direct consequence of its two dimensional (2D) lattice with underlying triangular symmetry, have attracted a great interest in recent two decades \cite{Geim}.
Advanced methods such as atom-by-atom assembling \cite{Gomes}, optical trapping of ultracold atoms in crystals of standing light-waves \cite{Tarruell} and nanopatterning of 2D electron gas in semiconductors \cite{Singha}, make it possible to design and fabricate artificial honeycomb lattices or artificial graphene, which are a unique playground for investigation and manipulation of a wide class of systems displaying massless Dirac quasiparticles and topological phases.
To replicate in a tunable manner the massless Dirac fermion
physics authors of Ref.\ \cite{Scarabelli} used high resolution electron beam lithography and reactive
ion etching in order to construct artificial honeycomb lattices with periods as small as 50 nm in GaAs/AlGaAs quantum
wells hosting a 2D electron gas.
The lack of an energy bandgap in graphene or artificial graphene constrains their widespread application because the small band gap means a large off-current and a low on/off ratio. Many attempts have been made to create an energy gap between the conduction and the valence bands of graphene. Cutting graphene into nanoribbons \cite{Han}, application of strain on graphene \cite{Guinea,Gui,Pereira,Cocco},
hydrogenating graphene with a certain pattern \cite{Gao}, and growth
of graphene on various substrates \cite{Zhou,Giovannetti} are examples of such attempts just to mention few works in this field.

In artificial graphene composed of semiconductor quantum dots (QD) there are additional possibilities of band structure manipulation via variations of the QD shapes, sizes and external factors such as transverse magnetic and in-plane electric fields \cite{Mughnetsyan1,Mughnetsyan2}.

It is well-known that the description of the motion of an electron in a magnetic field is significantly modified when considering a periodic modulation of the electron's potential energy. This fact is connected with the commensurability conditions of the two characteristic length-scales describing the structure, namely, the magnetic length and the lattice constant. It has been shown by Azbel \cite{Azbel} and Hofstadter \cite{Hofstadter} that the original unit cell (UC) of the superlattice (SL) can not describe its translational periodicity when a homogeneous transverse magnetic field is applied. In this case one has to introduce so called magnetic UC which simultaneously contains an integer number of magnetic flux quanta and UCs of the original lattice. As a result, the energy spectrum of an electron displays a fractal structure known as the ``Hofstadter's butterfly", that has been obtained theoretically \cite{Hofstadter, Guillement, Beugeling, Gumbs1, Rokaj} as well as observed experimentally \cite{Yang, Dean}. 

The description of the electron motion in graphene subjected to a transverse homogeneous magnetic field is usually based on the Peierls substitution in tight binding models, or the Dirac Hamiltonian \cite{Goerbig}. This approach relies on the assumption that the magnetic field effects on the tunneling of an electron through the sites of the graphene lattice only by means of the addition of corresponding magnetic phases in the hopping parameters. The Dirac Hamiltonian is applicable when there is only one conducting electron in each site of the lattice leading to the emerging of relativistic electrons near the band's touching points. These assumptions being well justified for graphene, are not so for artificial graphene-like semiconductor structures. For more complete description of the 2D electron's motion in such artificial systems with taking into account the effect of the magnetic field on the degree of the confinement of electron in each QD as well as on the magnitudes of the hopping parameters we develop our theoretical study in the frame of the basis functions proposed initially by Ferrari \cite{Ferrari} and used thereafter by several authors for calculations of band structure and magneto-optical properties of modulated 2D electrons \cite{Silberbauer,Gudmundsson1,Gudmundsson2,Gudmundsson3}.

Based on this method we have developed in the present paper a comparative study of the electronic states and the magnetization of the honeycomb artificial graphene-like lattices composed of cylindrical and elliptical QDs to explore the effect of the structure symmetry-breaking on the measurable equilibrium properties. Our calculations indicate on a topological change in the miniband structure as well as qualitative modifications to the Hofstadter spectrum, and the magnetization of the honeycomb SL due to the ellipticity of QDs. 

The paper is organized as follows: section \ref{theory} is devoted to the description of the theoretical model, in section \ref{discussion} the obtained results are discussed, and finally, in section \ref{conclusions} the conclusions are presented. 

\section{THEORY}
\label{theory}
\begin{figure}
    \centerline{\includegraphics[width=0.4\textwidth]{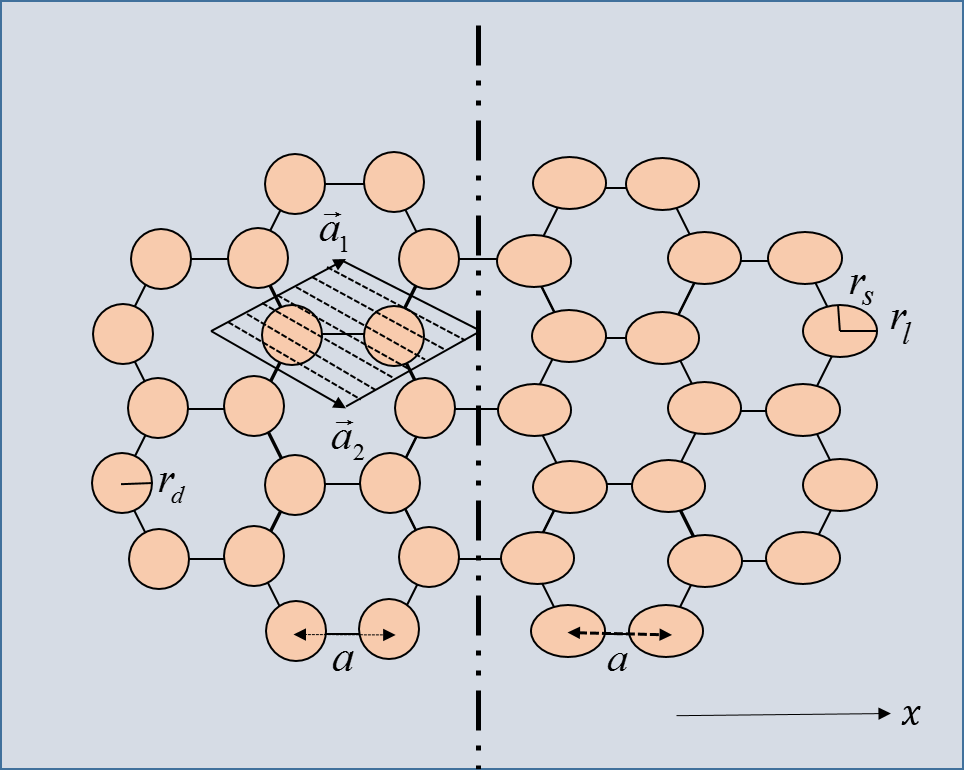}}
\caption{(Colour on-line) The schematic view of the 2D SLs of circular (the left half of the figure) and elliptical (the right half of the figure) QDs. On the figure $a$ is the distance between two nearest QDs, while $\vec{a}_{1}$ and $\vec{a}_{2}$ are the lattice vectors. $r_{d}$ is the radius of circular QDs, while $r_{s}$ and $r_{l}$ are the small and the large semiaxes of elliptical QDs, respectively.}
\label{SL}
\end{figure}

Let us consider a 2D lattice composed of planar QDs exposed to a transverse homogeneous magnetic field with induction $\mathbf{B}=\bm{\hat{e}}_{z}B$, where $\bm{\hat{e}}_{z}$ stands for the unit vector in the direction perpendicular to the lattice plane. The spinless one-electron Hamiltonian of such a system in the effective mass approxiamtion is
\begin{equation}
    H=H_{0}+V(\mathbf{r}),
\end{equation}
where 
\begin{equation}
    H_{0}=\frac{1}{2m}\left(\mathbf{p}+\frac{e\mathbf{A}}{c}\right)^{2},
\end{equation}
and $V(\mathbf{r})=V(\mathbf{r}+n_{1}\mathbf{a}_{1}+n_{2} \mathbf{a}_{2})$ is the periodic potential of the SL with lattice vectors $\mathbf{a}_1$ and $\mathbf{a}_2$, $n_{1}$ and $n_{2}$ are integers, $\mathbf{p}=-i\hbar\bm{\nabla}$ is the momentum operator, $\hbar$ is the reduced Planck's constant, $m$ is the effective mass, and $c$ the speed of light. We assume that the SL consists of circular or elliptical QDs (see Fig.\ \ref{SL}) with a rectangular potential profile. Namely, $v(\mathbf{r})=0$ inside each QD and $v(\mathbf{r})=v_{0}$ in the surrounding medium. Note, that both SLs with circular and elliptical QDs have the same translation vectors and all the elliptical QDs have the spatial orientation along the ``$x$" axis. Using the symmetric gauge for the vector potential $\mathbf{A}=(B/2)(-y,x)$ the Hamiltonian (2) reads as
\begin{equation}
  H_{0}=\frac{\hbar ^2}{2m}{\left (\left( -i \frac{\partial}{\partial x}-\frac{y}{2l_{B}^2}\right)^2+\left(-i \frac{\partial}{\partial y}+\frac{x}{2l_{B}^2} \right)^2\right)},
\end{equation}
where $l_{B}=(c\hbar/e B)^{1/2}$  is the magnetic length.
The eigenfunctions of the Hamiltonian (3) are
\begin{equation}
    \varphi_{n_{L}}(r)=\frac{1}{\sqrt{2\pi l_{B}^{2} n_{L}!}}\left(\frac{x+iy}{\sqrt{2} l_{B}}\right)^{n_{L}} e^{- \frac{r^2}{4l_{B}^{2}}},
\end{equation}
where $n_{L}$ indicates the corresponding Landau level.

It is well known that the translation operator $T(\mathbf{R})=\exp{(i\mathbf{R} \mathbf{p}/\hbar)}$ with $\mathbf{R}=n_{1}\mathbf{a}_{1}+n_{2}\mathbf{a}_{2}$ does not commute with the Hamiltonian (2). Instead, the so-called magnetotranslation operator $S(\mathbf{R})=\exp{((i e/\hbar c)\mathbf{A}(\mathbf{R})\mathbf{r})}T(\mathbf{R})=\exp{((i/2l_{B}^2)(\mathbf{R}\times{\mathbf{r}})\hat{e}_{z})} T(\mathbf{R})$ which commutes with the Hamiltonian (2) can be used for construction of a complete and orthogonal set of basis functions for description of the motion of an electron with the Hamiltonian (1). On the other hand, magnetotranslation operators for any two lattice vectors $\mathbf{R_{1}}$ and $\mathbf{R_{2}}$ commute in the only case when there is an integer number of magnetic flux quanta in the area $|\mathbf{R_{1}}\times \mathbf{R_{2}}|$
\begin{equation}
    [S(\mathbf{R}_{1}),S(\mathbf{R}_{2})]=0, \quad if \quad  |\mathbf{R}_{1} \times \mathbf{R}_{2} | = 2 \pi u l_{B}^{2}
\end{equation}
where $u$ is an integer.

If one expresses the magnetic flux per unit cell of the SL as $\Phi/\Phi_{0}=pq/h_{1}h_{2}$,
where $p,q,h_{1}$ and $h_{2}$ are integers, the vectors satisfying the condition (5) will be related with original lattice vectors as follows: $\mathbf{R}_{1}=h_{1}\mathbf{a}_{1}$ and   $\mathbf{R}_{2}=h_{2}\mathbf{a}_{2}$.
As is shown in Ref.\ \cite{Ferrari} a complete set of basis functions can be constructed out using the primitive magnetotranslations $S(\mathbf{c})$ and $S(\mathbf{d})$ with $\mathbf{c}=\mathbf{R}_{1}/p$ and $\mathbf{d}=\mathbf{R}_{2}/q$. Taking into account the conditions of periodicity
\begin{equation}
    S(\mathbf{R}_{1}) \phi= e^{i \theta _{1}} \phi, \qquad  S(\mathbf{R}_{2}) \phi= e^{i\theta_{2}} \phi
\end{equation}
the basis functions can be expressed as
\begin{equation}
\begin{split}
     &\phi^{n_{1},n_{2}}_{n_{L}} (r)=\\
     &(pq)^{-1/2} \sum_{m,n=-\infty}^{\infty} [S(\mathbf{c}) e^{-i \mu}]^m  [S(\mathbf{d}) e^{-i \nu}]^n \phi_{n_{L}}(\mathbf{r}),
     \end{split}
 \end{equation}
where
 \begin{equation}  
  \begin{array}{l}
    \mu = (1/p) (\theta_{1}+2 \pi n_{1}) , \quad \quad \quad n_{1}={0,...,p-1},\\

    \nu=  (1/q) (\theta_{2}+2 \pi n_{2}) , \quad \quad \quad n_{1}={0,...,p-1}.\\
  
  \end{array}
 \end{equation}
In the absence of magnetic field, $\theta_{1}$ and $\theta_{2}$ are proportional to the components of the wave vector in the SL.
It has been shown that the norm of the wave function (7), is nonzero when $(\mu,\nu)\neq (\pi,\pi)$ and can be expressed as \cite{Ferrari, Silberbauer}
 \begin{equation}
     \parallel  \phi^{n_{1},n_{2}}_{n_{L}} \parallel =\sum_{m,n=-\infty}^{\infty} (-1)^ {mn} e^{i(\mu m+ \nu n)} e^{-\frac{\mid  nc+md \mid ^{2}}{4l_{B}^{2}}}. 
 \end{equation}

A periodic SL potential can be expanded in a Fourier series 
\begin{equation}
V(\mathbf{r})= \sum_{\mathbf{G}} v(\mathbf{G})e^{i\mathbf{G}\mathbf{r}},
\end{equation}
where $\mathbf{G}= G_{1} \mathbf{g}_{1}+G_{2} \mathbf{g}_{2}$ are the reciprocal lattice vectors with site-vectors $\mathbf{g}_{1}$ and $\mathbf{g}_{2}$, and integers $G_{1}$ and $G_{2}$. 
For the periodic array composed of cylindrical or elliptical QDs, respectively
\begin{eqnarray}
   v(\mathbf{G})_{\text{cyl}}= 
    \left\{
    \begin{array}{l}
    \frac{v_{0}}{s_{0}} 2 \pi r_{d}^{2} \hspace{0.2cm} \text{if} \hspace{0.2cm} G_{1}=G_{2}=0; \vspace{0.5cm} \\ 
    \frac{v_{0}}{s_{0}}  e^{-i\frac{2\pi}{3}(G_{1}+G_{2})}(1+e^{-i\frac{2\pi}{3}(G_{1}+G_{2})}) \times \\
   \frac{3r_{d}a}{\sqrt{(G_{1}+G_{2})^{2}+3(G_{1}-G_{2})^{2}} } \times \\ J_{1}\left(\frac{2\pi r_{d}\sqrt{(G_{1}+G_{2})^{2}+3(G_{1}-G_{2})^{2}} }{3a}\right), \\ 
    \text{otherwise}
  \end{array}
\right.
\end{eqnarray}
and 
 \begin{equation}
\begin{split}
   &v(\mathbf{G})_{\text{el}}=\frac{v_{0}}{s_{0}}  e^{-i\frac{2\pi}{3}(G_{1}+G_{2})}
   \left(1+e^{-i\frac{2\pi}{3}(G_{1}+G_{2})}\right) \times \\
   &\int \limits_{0}^{r_{e}} \int \limits_{0}^{2\pi} e^{-i\frac{2\pi}{3a}r\left((cos\phi+\sqrt{3} sin\phi)G_{1}+(cos\phi-\sqrt{3} sin\phi)G_{2}\right)} r dr d\phi,
 \end{split}
\end{equation} 
 where 
 \begin{equation}
   r_{e}=\frac{r_{s}r_{l}}{\sqrt{{r_{s}^{2} \cos^{2}\phi+r_{l}^{2} \sin^{2}\phi}}},  
 \end{equation}
$r_{s(l)}$ is the small(large) semi-axis of the elliptical QD, $a$ is the distance between the nearest QDs and $s_{0}$ is the area of the SL unite cell. Now, the calculation of the potential matrix elements will simply be reduced to the calculation of the ones for the exponent in Eq.\ (10). These matrix elements are not zero when the following conditions are fullfilled
\begin{equation}  
  \begin{array}{l}
    G_{1}h_{1}+n_{1}-n_{1}'=Mp\\

    G_{2}h_{2}+n_{2}-n_{2}'=Nq,\\
  
  \end{array}
 \end{equation}
 with $M$ and $N$ integers, and can be expressed as
\begin{equation}
\begin{split}
   & \langle n_{1}',n_{2}',n_{L}'\mid  e^{i\mathbf{G}\mathbf{r}} \mid   n_{1},n_{2},n_{L} \rangle =\\ 
  &\frac{Y(G)^{n_{L}',n_{L}} 
 {T^{n_{1}',n_{2}'}_{n_{1},n_{2}}}(G) \exp{(-|G|^{2}/4l_{B}^{2})}}{\| \phi ^ {n_{1}',n_{1}'}_{n_{L}'} \|  \| \phi ^ {n_{1},n_{1}}_{n_{L}} \|},
 \end{split}
\end{equation} 
 where
\begin{equation}
\begin{split}
&T^{n_{1}',n_{2}'}_{n_{1},n_{2}}(G)=\\
&\sum_{\Lambda , \Omega =- \infty}^{\infty} (-1)^{\Lambda \Omega} e^{i (\mu' \Lambda + \nu' \Omega)} {e^{-(i/2)G(\Lambda c + \Omega d)}} ^{\ast} \times \\ 
&\exp{(-(1/4l_{B}^{2}) |\Lambda c+\Omega d|^{2})},
\end{split}
\end{equation}
\begin{eqnarray}
Y^{m,n}(G)= 
    \\ \nonumber
    \left\{
  \begin{array}{l}
\sqrt{\frac{m!}{n!}} e^{(-1/4)|G|^{2}}(iG^{\ast}/\sqrt{2})^{n-m} L^{n-m}_{m}(|G|^{2}/2), \\ \nonumber
  \quad n\geq m,\\
     \sqrt{\frac{n!}{m!}} e^{(-1/4)|G|^{2}}(iG/\sqrt{2})^{m-n} L^{m-n}_{n}(|G|^{2}/2), \\ \nonumber
    \quad m\geq n,\\
    
  \end{array}
\right.
\end{eqnarray}
and $L^{\beta}_{\alpha}(x)$ are the Laguerre polynomials \cite{Silberbauer}.
In the right-hand side of Eqs.\ (15), (16) and (17) a complex notation for the vectors has been used $(G=G_{x}+iG_{y}$,  $c=c_{x}+ic_{y}$, $d=d_{x}+id_{y})$.

The density of states (DOS) is defined as
\begin{equation}
    \rho(E)=\frac{1}{S} \sum_{i} \delta(E-E_{i}),
\end{equation}
where the summation is carried out over all the quantum states and $S$ is the area of the sample. Considering the quasi-continuous energy spectrum inside each miniband, Eq.\ (18) can be transformed as
\begin{equation}
    \rho(E)=\frac{1}{(2\pi)^{2}S}\sum_{j} \int_\mathrm{FBZ} d\theta_{1}d\theta_{2} \delta \left(E-E_{j}(\theta_{1},\theta_{2})\right),
\end{equation}
 where the integration is carried out over the FBZ and $j$ is the number of a miniband. In numerical calculation we have replaced the Dirac delta function $\delta$ by a Lorentzian function with small energy width $\Gamma=10^{-3}$ meV.

We calculate the orbital magnetization as
\begin{equation}
    \mathcal{M}=\frac{1}{(2\pi)^2}\sum_{j} \int_\mathrm{FBZ} d\theta_{1}d\theta_{2} f_{B}(E_{j}(\theta_{1},\theta_{2}))\mathcal{M}_{j}(\theta_{1},\theta_{2}),
\end{equation}
where $f_{B}(E)$ is the Fermi function with energy $E$, and the magnetization for each point in the reciprocal space is
\begin{equation}
    \mathcal{M}_{i}(\theta_{1},\theta_{2})= \frac{1}{2 c S} \int_{S} d \mathbf{r}\: (\mathbf{r} \times \mathbf{j}_{i,\theta_{1},\theta_{2}}(\mathbf{r})) \hat{e}_{z}
\end{equation}
 with
\begin{equation}
    \mathbf{j}(\mathbf{r})=-\frac{e}{2}(\hat{\mathbf{v}} | \psi(r)\rangle \langle \psi(r)|+| \psi(r)\rangle \langle \psi(r)|\hat{\mathbf{v}}),
\end{equation}
the current density operator.
The velocity operator is $\hat{\mathbf{v}}=(\hat{\mathbf{p}}+(e/c)A(\mathbf{r}))/m$.
Note, that the use of the formula (20), which is different from one we have previously used in \cite{Mansoury}, is connected with the discrete integer values of magnetic flux which one has to consider in a Hofstadter-like problem. 
\begin{figure}
     \centering
     \begin{subfigure}
         \centering
         \includegraphics[width=0.35\textwidth]{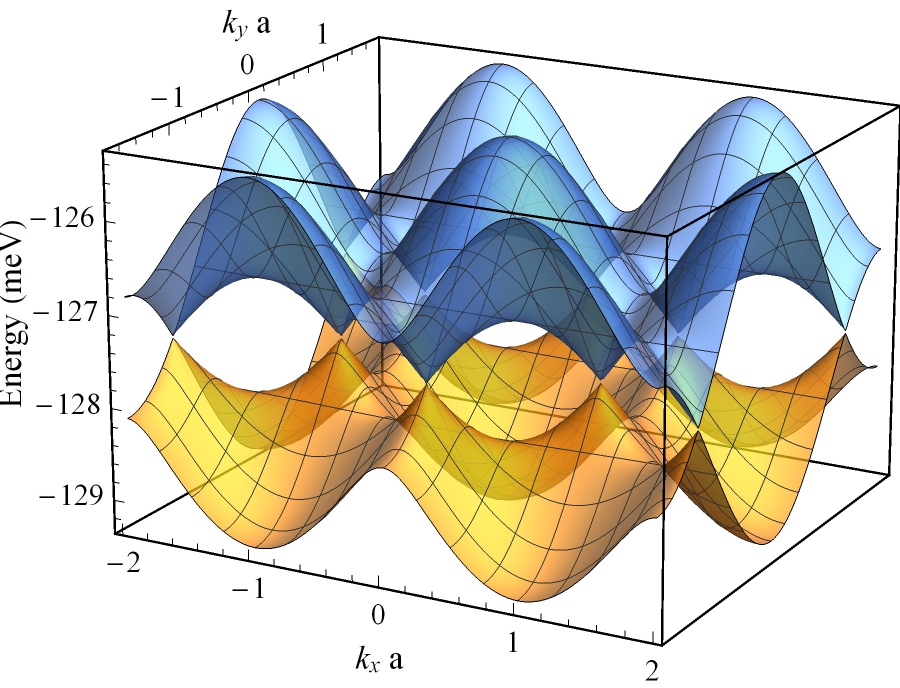}
        %\caption{$y=x$}
         %\label{HofnL5Cil}
     \end{subfigure}
     \vfill
     \begin{subfigure}
         \centering
         \includegraphics[width=0.35\textwidth]{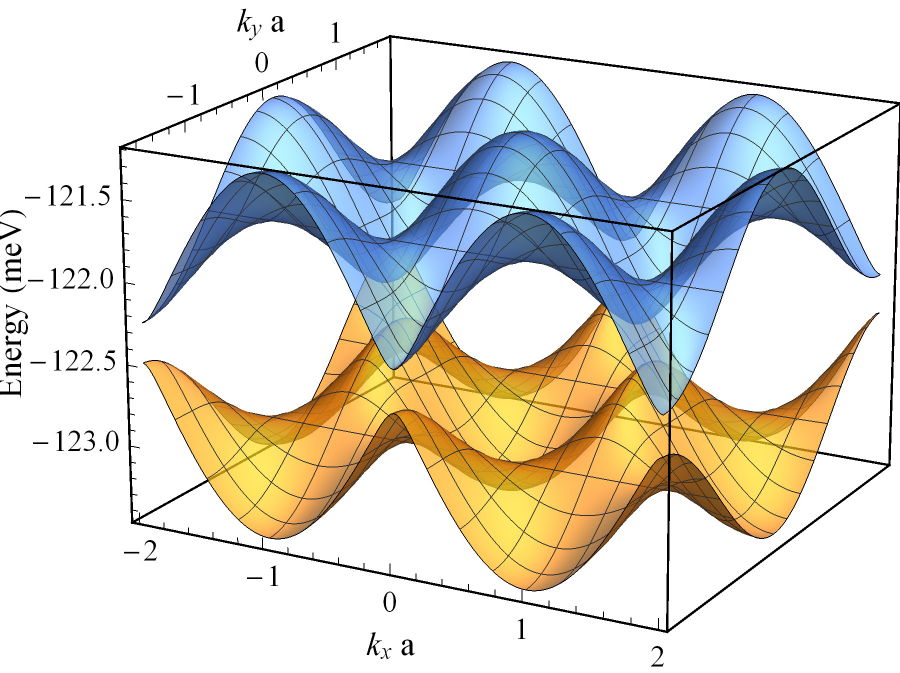}
        %\caption{$y=3sinx$}
          %\label{HofnL5Ellipt}
     \end{subfigure}
\caption{(Colour online) Electron energy dispersion surfaces in the absence of external magnetic field for a SL composed of circular (upper figure) and a SL composed of elliptical (lower figure) QDs.}
\label{dispB0}
\end{figure}
\begin{figure}
 \centering
\includegraphics[width=0.50\textwidth]{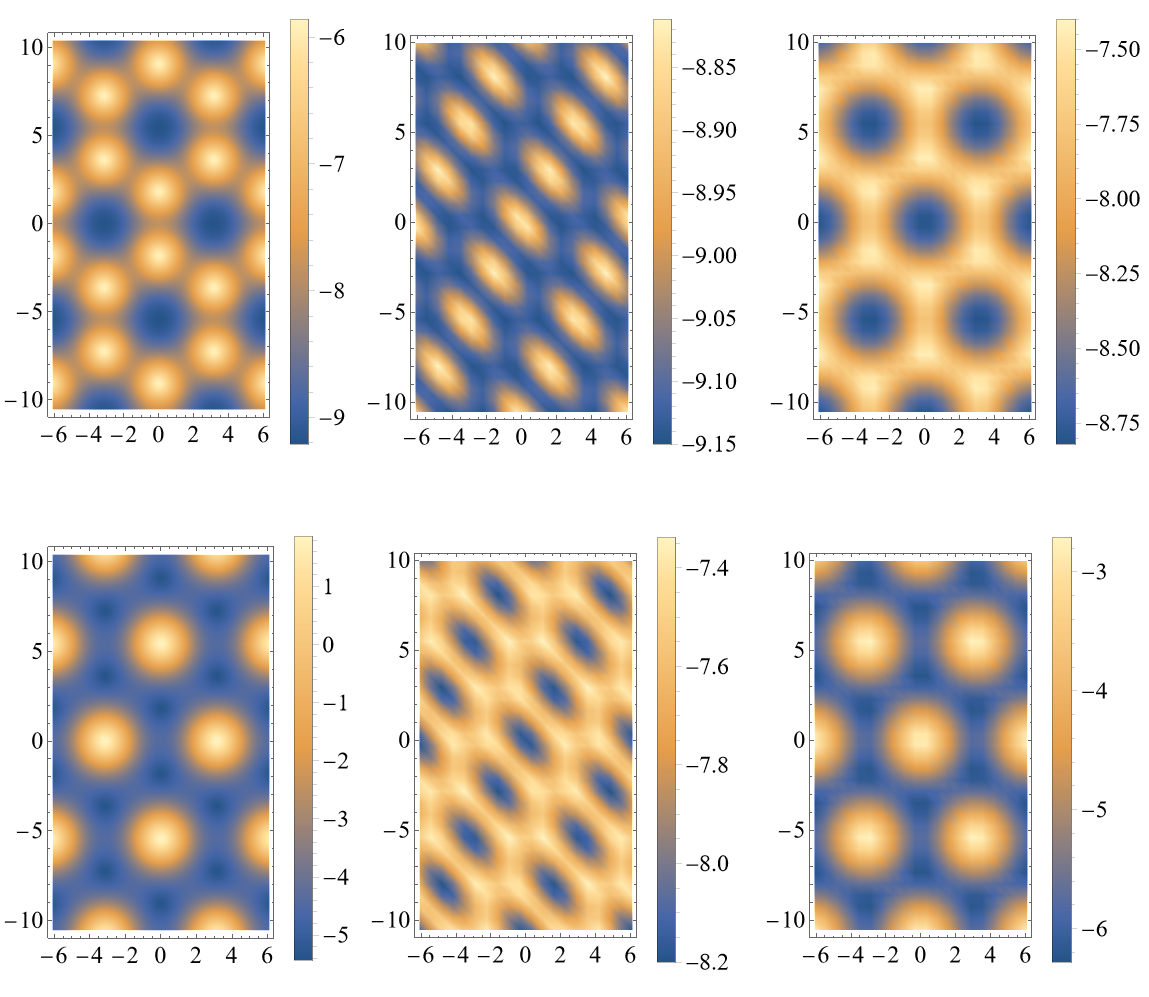}
\caption{(Colour online) Dispersion surfaces for the first (upper row) and the second (lower row) miniband in honeycomb SL of circular QDs. The horizontal axis in each panel is for $\theta_{x}=(1/2)(\theta_{1}+\theta_{2})$ and the vertical axis is for $\theta_{y}=(\sqrt{3}/2)(\theta_{1}-\theta_{2})$. The magnetic flux in the unites of flux quantum is $1$, $3/2$ and $2$ for the left, middle and the right columns, respectively. The values of energy are expressed in meV.}
\label{DispCirc}
 \end{figure}
 \begin{figure}
\includegraphics[width=0.50\textwidth]{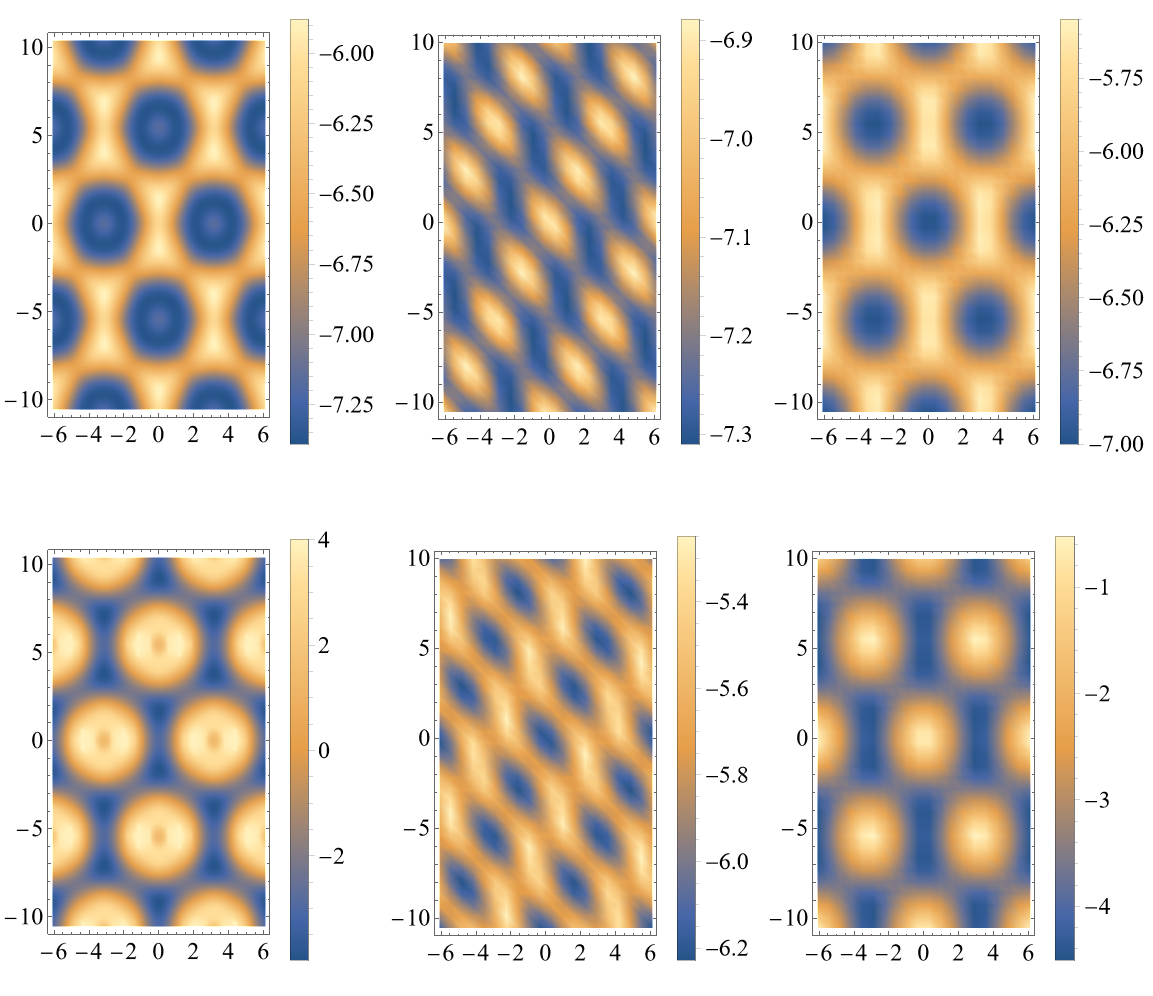}
\caption{(Colour on-line) Dispersion surfaces for the first (upper row) and the second (lower row) minibands in honeycomb SL of elliptical QDs. The horizontal axis in each panel is for $\theta_{x}=(1/2)(\theta_{1}+\theta_{2})$ and the vertical axis is for $\theta_{y}=(\sqrt{3}/2)(\theta_{1}-\theta_{2})$. The magnetic flux in the unites of flux quantum is $1$, $3/2$ and $2$ for the left, middle and the right columns, respectively.}
\label{DispEllipt}
 \end{figure}

\begin{figure}
     \centering
     \begin{subfigure}
         \centering
         \includegraphics[width=0.35\textwidth]{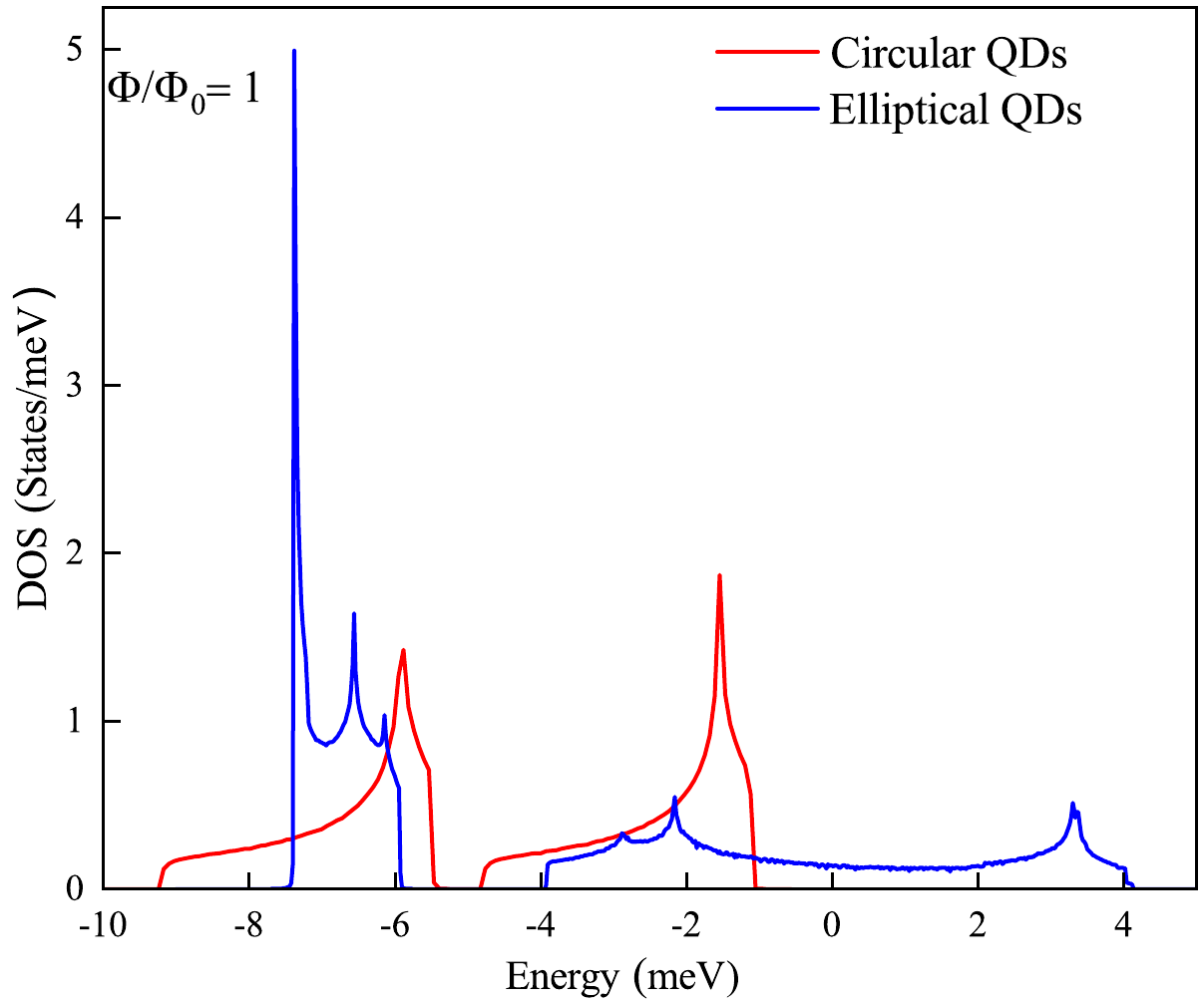}
        % \caption{$y=x$}
         \label{DOSa}
     \end{subfigure}
     \vfill
     \vspace{2mm}
     \begin{subfigure}
         \centering
         \includegraphics[width=0.35\textwidth]{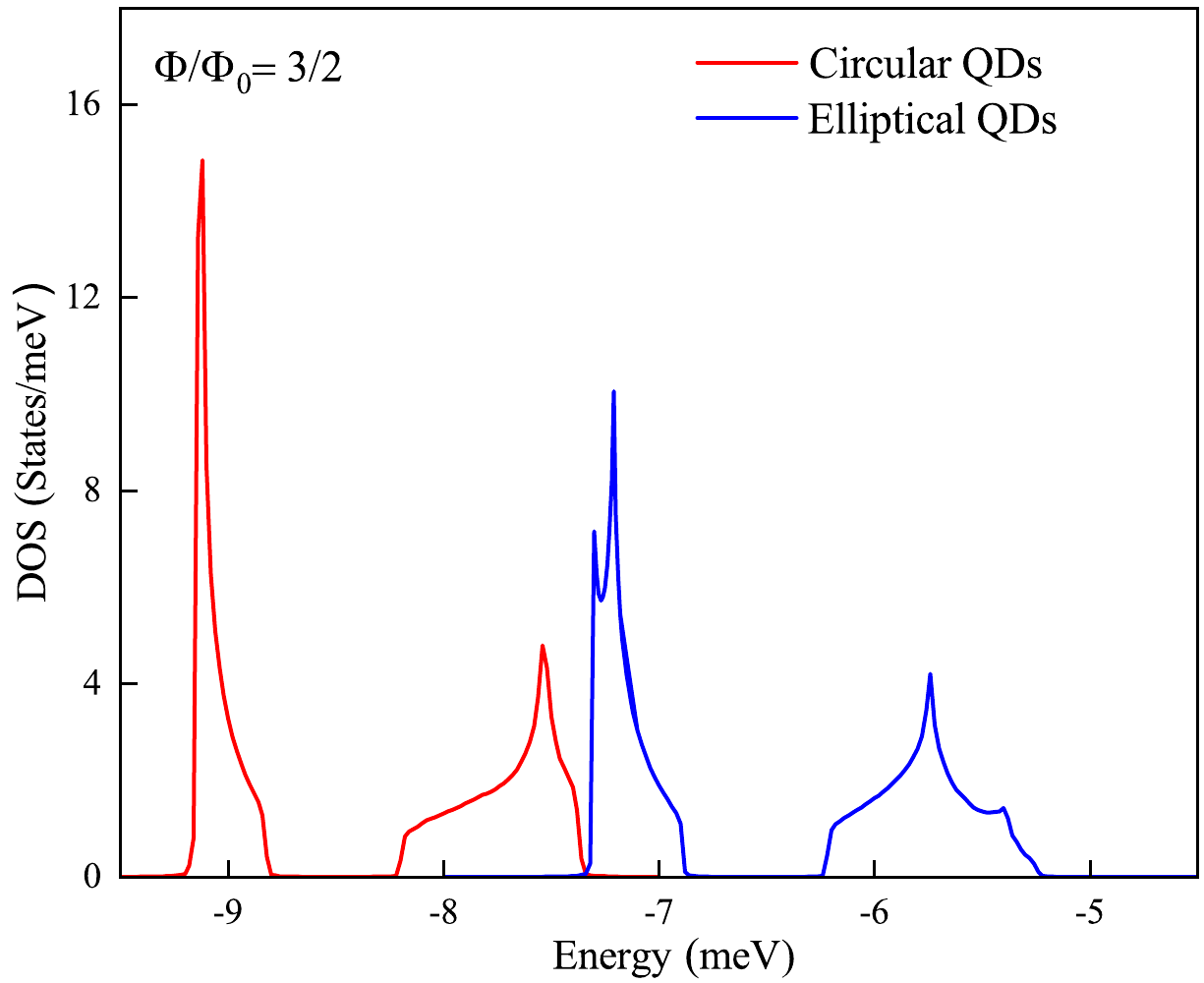}
        % \caption{$y=3sinx$}
          \label{DOSb}
     \end{subfigure}
    
     \vfill
     \vspace{2mm}
     \begin{subfigure}
         \centering
         \includegraphics[width=0.35\textwidth]{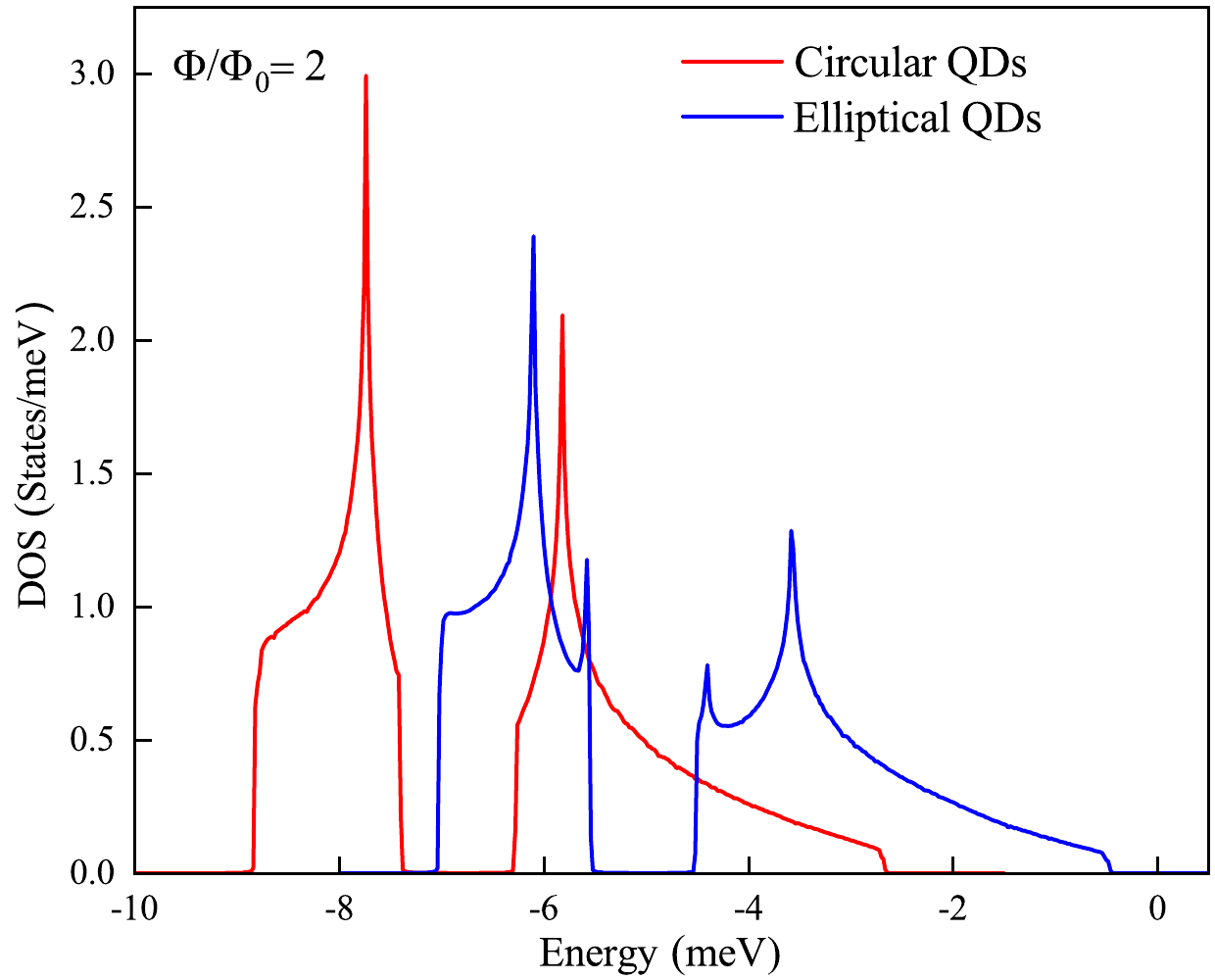}
        % \caption{$y=3sinx$}
          \label{DOSc}
     \end{subfigure}
     \vfill
     \vspace{2mm}
     \begin{subfigure}
         \centering
         \includegraphics[width=0.35\textwidth]{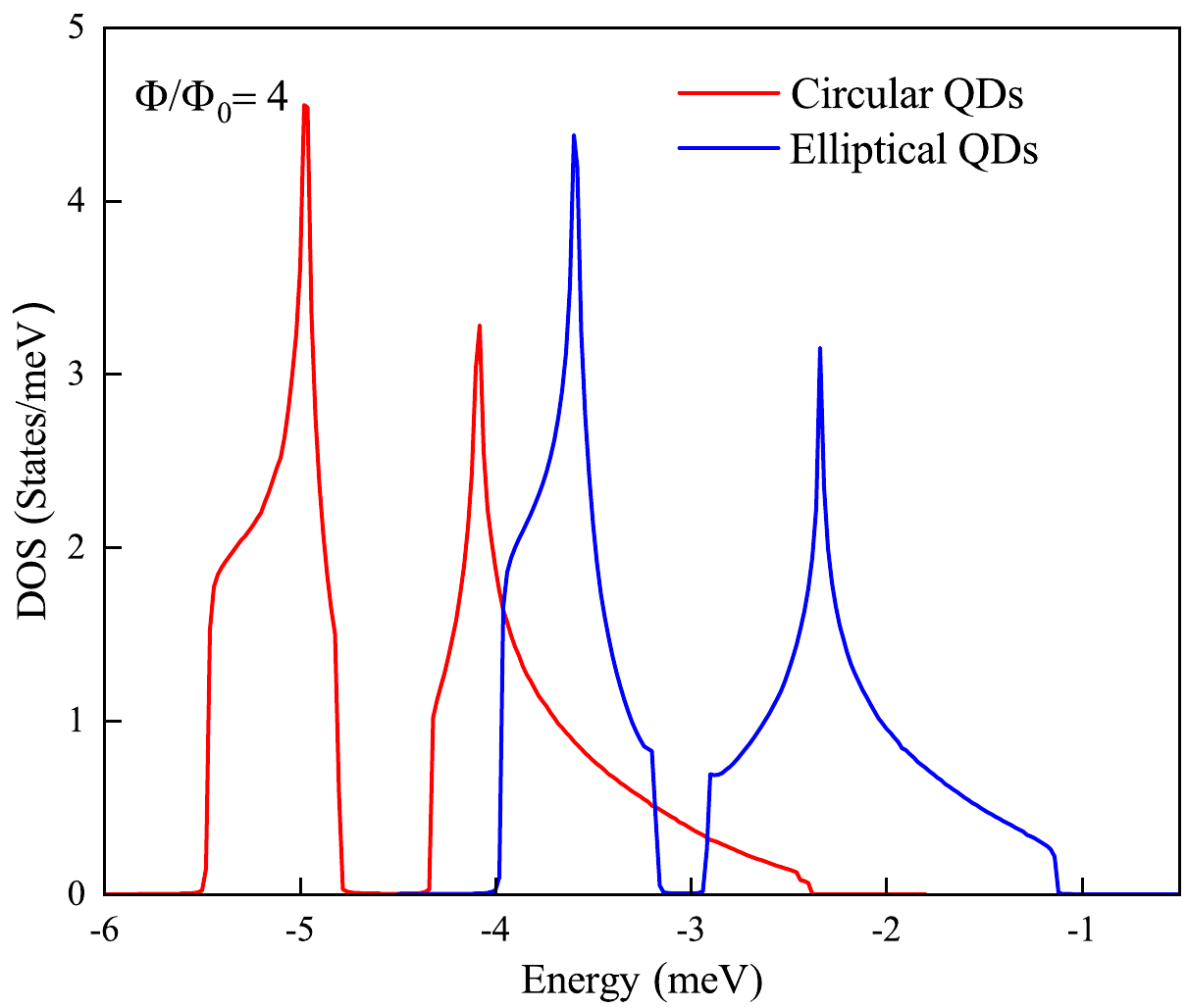}
        % \caption{$y=3sinx$}
          \label{DOSd}
     \end{subfigure}
    
        \caption{(Colour online) Density of states as a function of electron energy in honeycomb SLs composed of circular (red lines) and elliptical (blue lines) QDs.}
        \label{DOS}
\end{figure}
%%%%%%%%%%%%%%%%%%%%%%%%%%%%%%%%%%%%%%%%%%%%%%%%%%%%%%%%%%%%%%%%%%%
\begin{figure}
      \includegraphics[width=0.55\textwidth,viewport= 400 100 2800 2100]{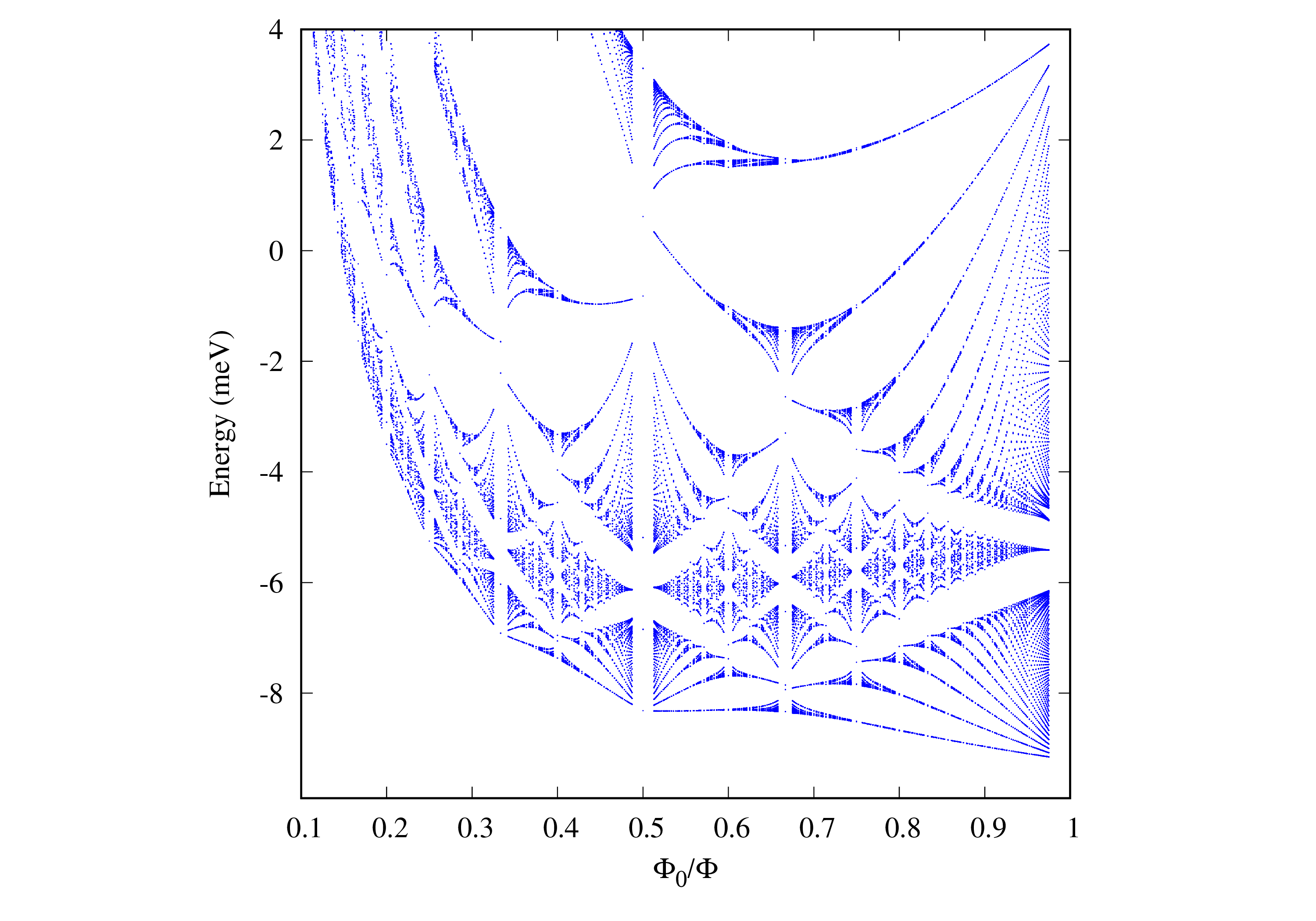}
      \includegraphics[width=0.55\textwidth,viewport= 400 400 2800 2100]{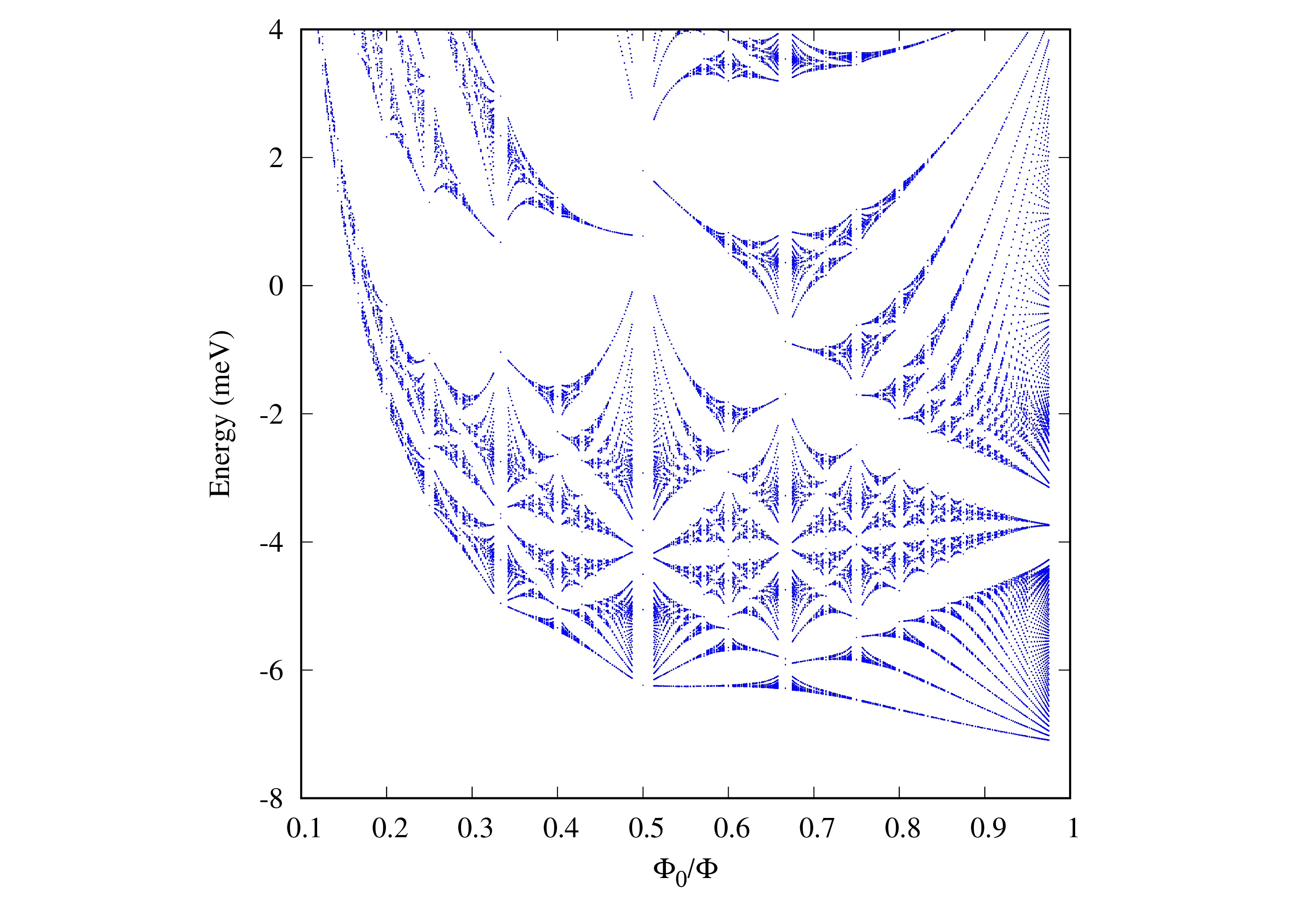}
      \vspace{1cm}
     \caption{(Colour online) Hofstadter spectrum for a honeycomb SL composed of circular (upper figure) and elliptical (lower figure) quantum dots for the case when two lowest Landau bands are considered in the expansion of the electronic wave function.}
    \label{Hof_nL1}
\end{figure}
\begin{figure}
      \includegraphics[width=0.55\textwidth,viewport= 400 100 2800 2100]{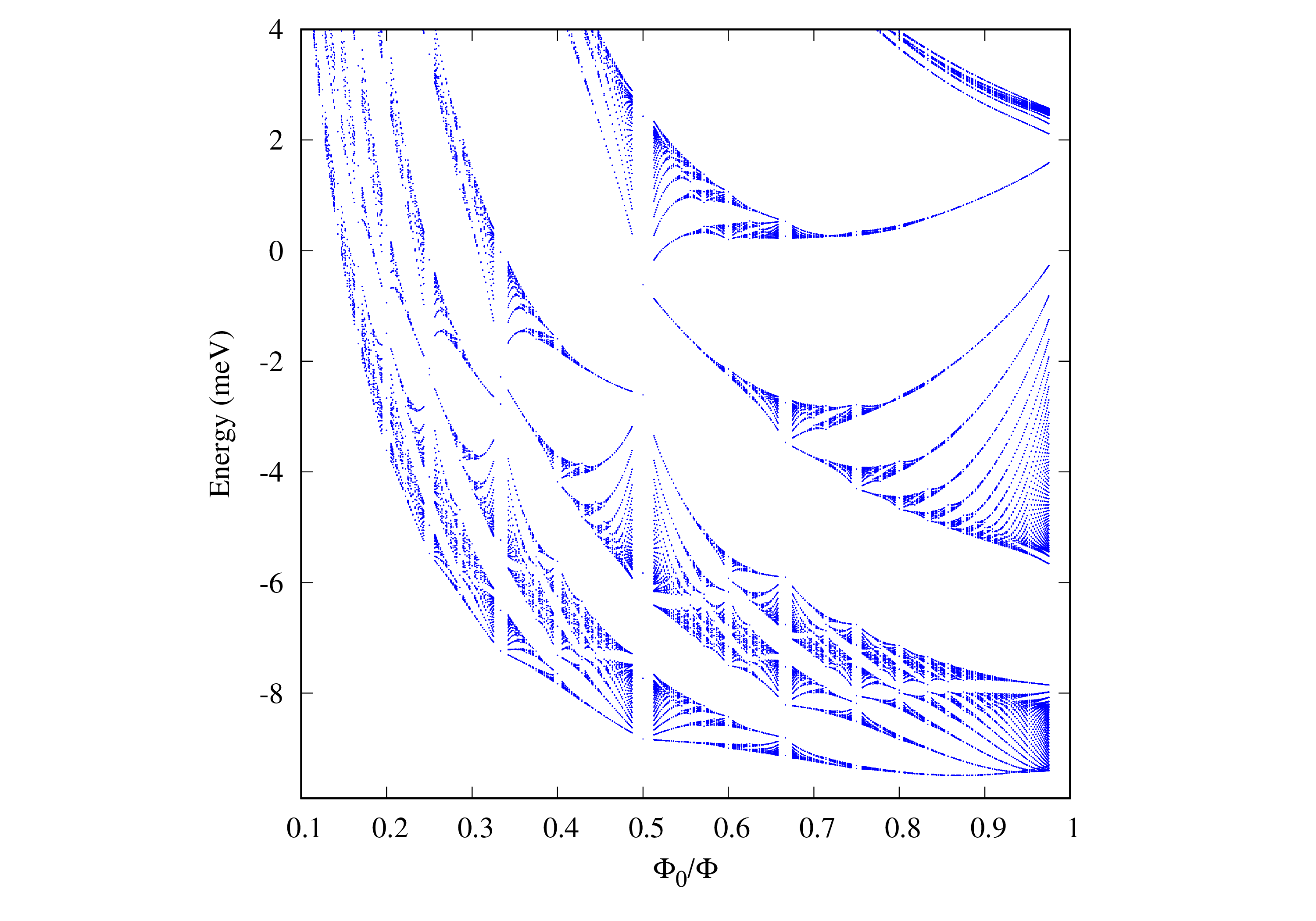}
      \includegraphics[width=0.55\textwidth,viewport= 400 400 2800 2100]{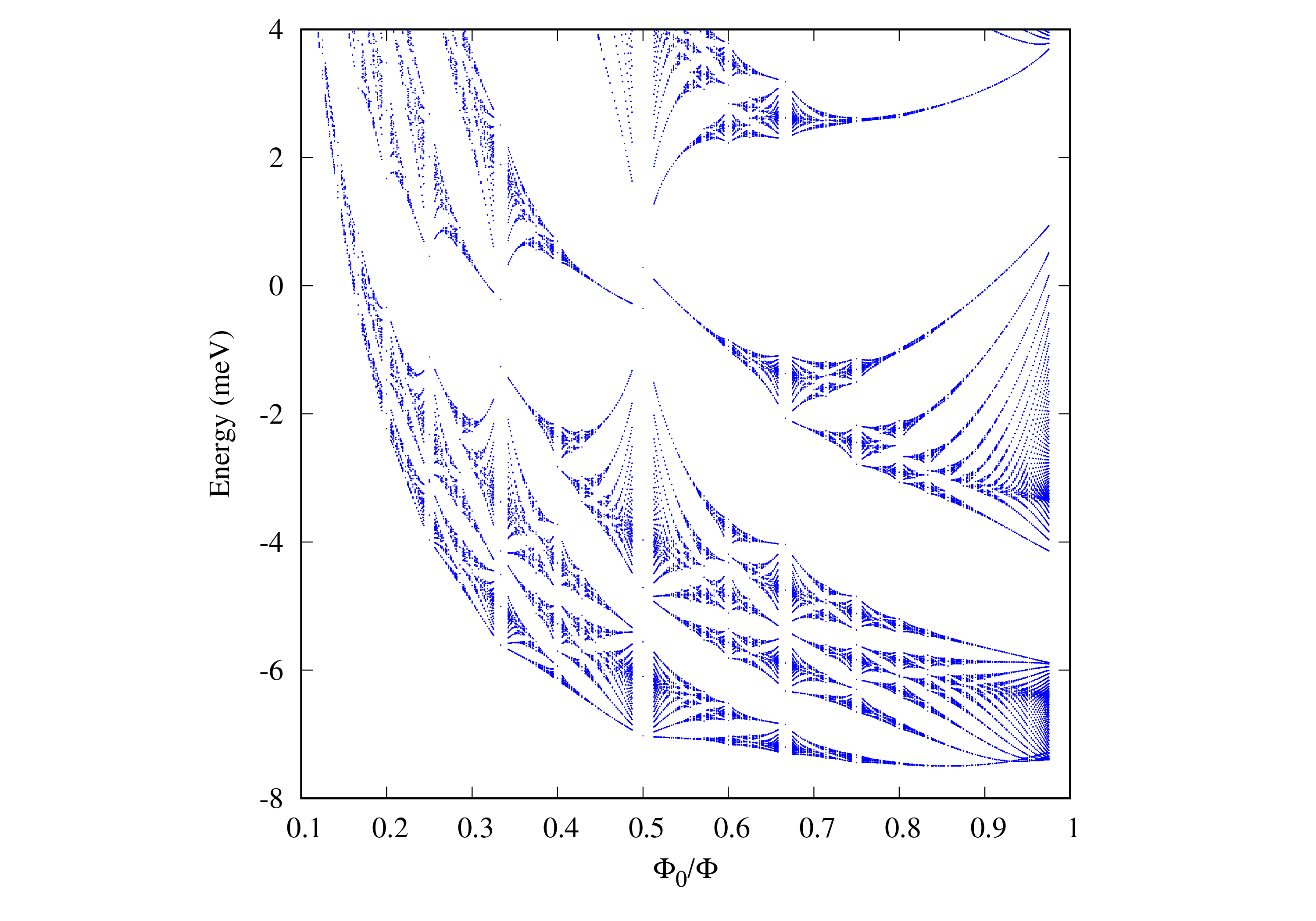}
      \vspace{1cm}
     \caption{(Colour online) Hofstadter spectrum for a honeycomb SL composed of circular (upper figure) and elliptical (lower figure) quantum dots for the case when six lowest Landau bands are considered in the expansion of the electronic wave function.}
    \label{Hof_nL5}
\end{figure}
\begin{figure}
     \centering
     \begin{subfigure}
         \centering
         \includegraphics[width=0.45\textwidth]{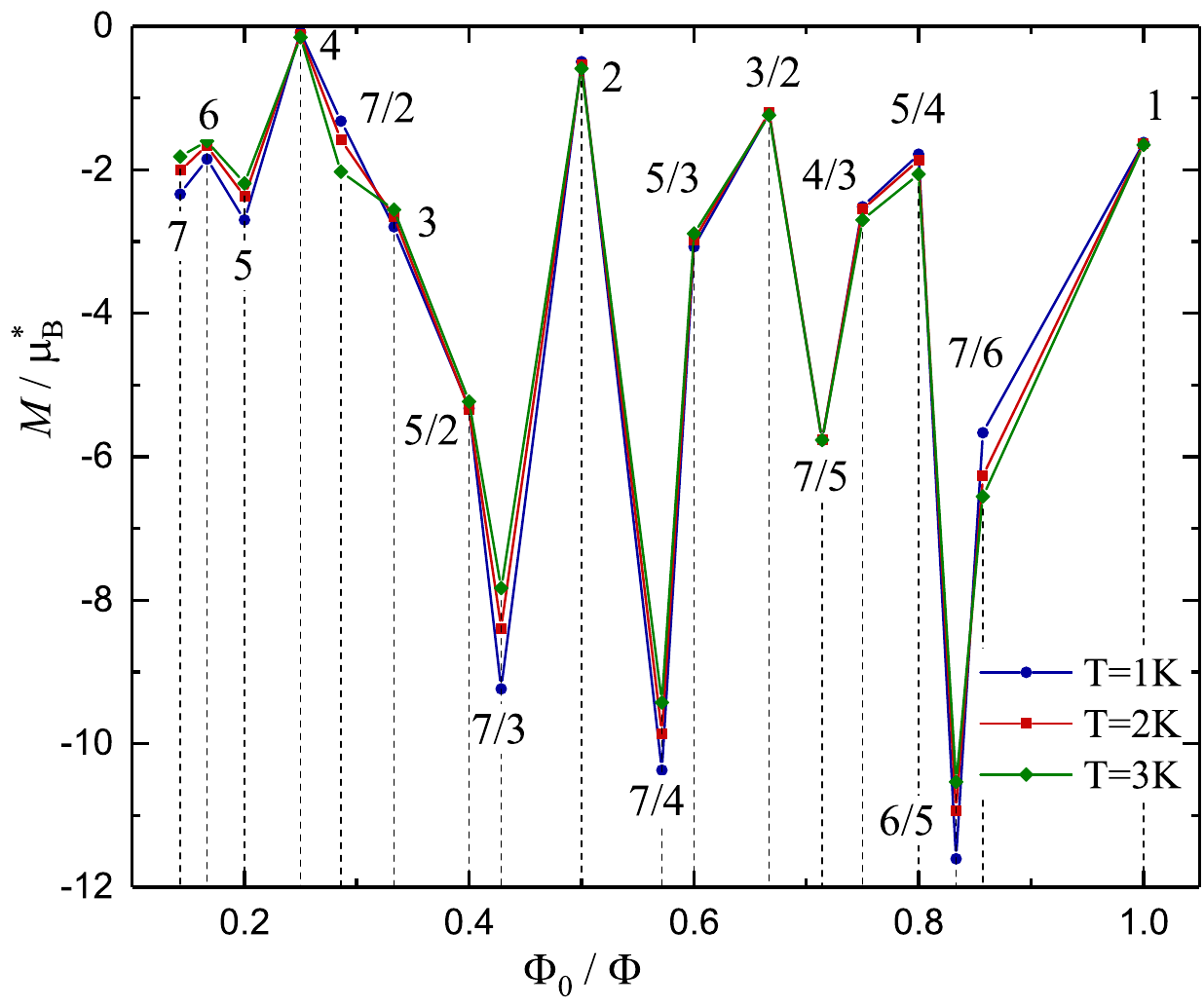}
        %\caption{$y=x$}
         %\label{HofnL5Cil}
     \end{subfigure}
     \vfill
     \begin{subfigure}
         \centering
         \includegraphics[width=0.45\textwidth]{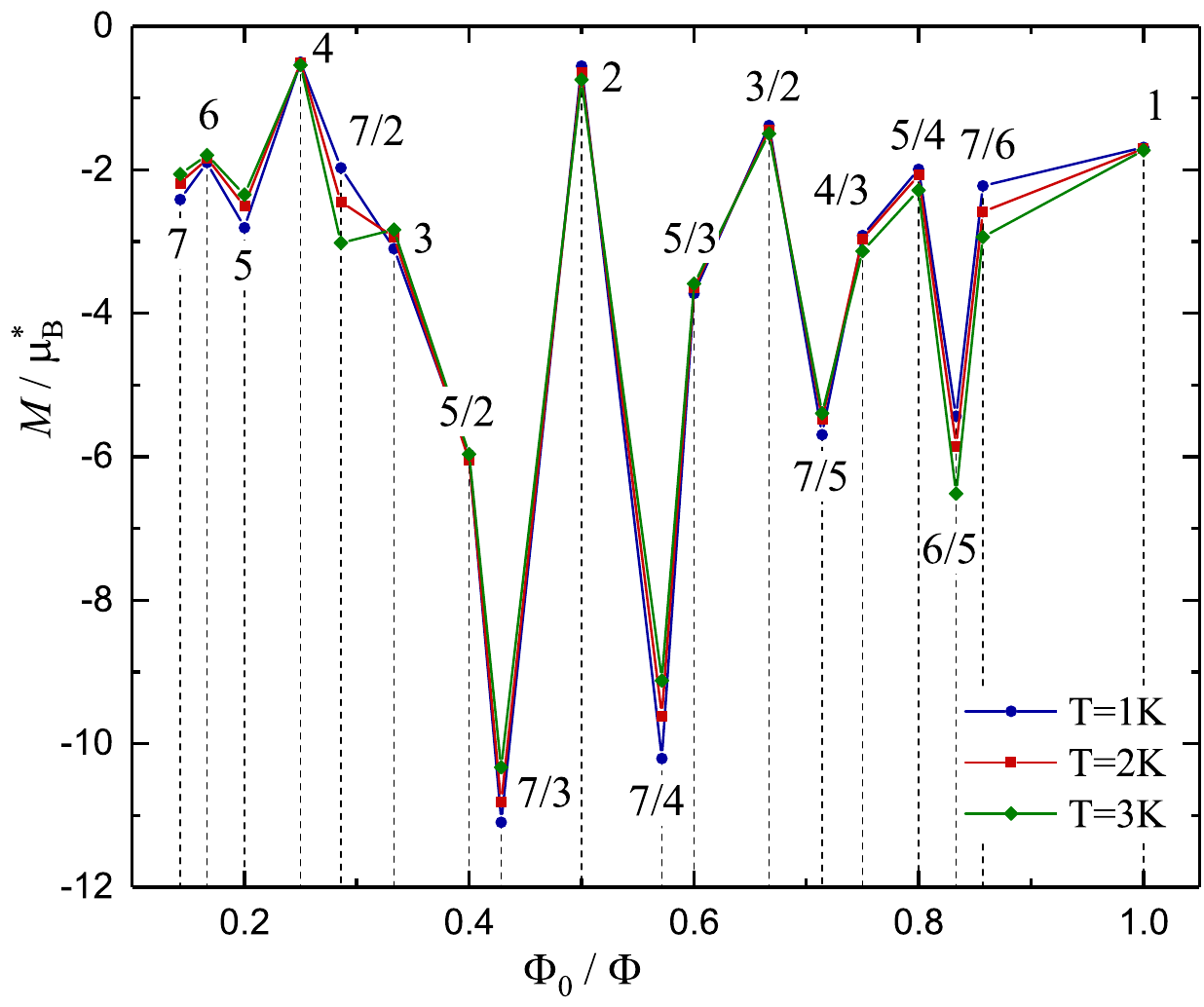}
        %\caption{$y=3sinx$}
         %\label{HofnL5Ellipt}
     \end{subfigure}
\caption{(Colour online) Orbital magnetization of honeycomb SL composed of circular (upper panel) and elliptical (lower panel) quantum dots for 18 rational values of the magnetic flux through the unit cell.}
\label{Magn}
\end{figure}
%%%%%%%%%%%%%%%%%%%%%%%%%%%%%%%%%%%%%%%%%%%%%%%%%%%%%%%%%%%%%%%%%%%%
\section{Discussion}
\label{discussion}
The numerical calculations are carried out for SLs composed of GaAs/Ga$_{1-x}$Al$_{x}$As QDs with the following values for the parameters: the radius of a circular QD $r_ {d}= 120 $ \r{A}, the small and large semiaxes of elliptical QD $r_{s}=0.8 r_{d}$ and $r_{l}= r_{d}$, respectively, the distance between two nearest QDs $a = 250$  \r{A}. When considering non-zero magnetic field we have chosen a shallow potential for each QD: $v_{0} = -16$ meV to have more clear picture of the magnetic field effect, while for the case of no magnetic field a value of the confining potential $v_{0}=-150$ meV is chosen. The electron effective mass $m=0.067m_0$, where $m_0$ is the free electron mass.

In Fig.\ \ref{dispB0} the dispersion surfaces for a SL composed of circular (the upper figure) and elliptical (lower figure) QDs are presented in the absence of external magnetic field. Here $k_{x}$ and $k_{y}$ are the Cartesian components of electron quasimomentum. It is obvious the qualitative coincidence of the dispersion surfaces for the SL of circular QDs with ones for graphene. As was expected, there is an energy gap between the two minibands for the SL composed of elliptical QDs. The gap opening and the topological change of the dispersion surfaces near the Dirac points are consequences of the triangular symmetry breaking of the system. Instead, the SL with the elliptical QDs reveals a rectangular symmetry which is expressed on the dispersion surfaces as well.

Fig.\ \ref{DispCirc} represents the density plots of the dispersion surfaces for a SL composed of circular QDs. The considered values of the magnetic flux per UC are $ \Phi / \Phi_{0} = 1$, $3/2$, and $2$, respectively for the left, the middle and the right columns in the figure. The upper row of the figure corresponds to the 1st, while the lower row is for the 2nd miniband. First of all it is obvious that the dispersion surfaces retain their triangular symmetry when there is an integer number of magnetic flux quanta per UC (the left and the right columns of the Fig.\ \ref{DispCirc}). However, the fractional number of flux quanta per UC leads to the destruction of the triangular symmetry. For $\Phi/\Phi_{0}=3/2$ one of the lattice vectors of the system with magnetic field is twice that of the corresponding lattice vector of the original lattice, which leads to the contraction of the FBZ in the perpendicular direction (see the middle column of Fig.\ \ref{DispCirc}). Another interesting phenomenon is the shift of the positions in the FBZ of maxima and minima of the dispersion surfaces corresponding to even and odd numbers of flux quanta per UC with regard to each other (compare the upper figure in Fig.\ \ref{dispB0} with the left and the right columns of Fig.\ \ref{DispCirc}). In all the cases with non-zero magnetic field a finite gap between the minibands is opened. This result is the consequence of the magnetic-phase interference between the states localized in two QDs in the same UC. 

Fig.\ \ref{DispEllipt} represents the same as Fig.\ \ref{DispCirc}, but for a SL composed of elliptical QDs.
In this case the SL has a rectangular symmetry, which is not destroyed by the magnetic field when there is an integer number of flux per UC. The energy values shown on the legends of the figure indicate on the decrease of the gap between the minibands with the increase of the magnetic flux. One can also observe that the rectangular symmetry is better expressed for larger integer numbers of magnetic flux per UC (compare the left and the right columns of the Fig.\ \ref{DispEllipt}).

The dependencies of the density of states on the electron energy for different values of magnetic flux quanta per UC is shown in Fig.\ \ref{DOS}. The red curves are plotted for a SL with circular and the blue ones are for a SL with elliptical QDs. As is expected, the minibands, and hence, the DOS are shifted to higher energies for elliptical QDs as the size-quantization in elliptical QDs is stronger. In all the cases there is a finite-length energy region where DOS is zero, which corresponds to the gap between the minibands. It is clear that the DOS which corresponds to circular QDs always have one maximum in each miniband. In contrast, the DOS for the SL with elliptical QDs has two obvious maxima in each miniband when $\Phi/\Phi_{0}=3/2$ or $2$. When $\Phi/\Phi_{0}=4$ the maxima near the energy gap are disappeared. For $\Phi/\Phi_{0}=1$ one can observe three maxima in each miniband. The change in the number of the DOS maxima significantly affects the optical characteristics of the system, which means that magnetic field can be used as an efficient tool for manipulations of the optical parameters of a honeycomb SL.

We demonstrate the fractal structure of the energy as a function of inverse magnetic flux per UC in Figs.\ \ref{Hof_nL1} and \ref{Hof_nL5}. In order to save the computational time, as well as to make figures more readable we present here the energies of an electron only for $\theta_{1,2}= \pm 0.99\pi, 0$. Fig.\ \ref{Hof_nL1} is obtained by using only two Landau bands with $n_{L}=0$ and $n_{L}=1$ in the expansion of the wave function by the basis functions (7), while the results shown in Fig.\ \ref{Hof_nL5} correspond to a basis with six Landau bands ($n_{L}=0 - 5$). Note, that the basis with six Landau bands provides results with high enough accuracy (the estimated relative error is around $1 - 2\%$). Nevertheless, we present here also the case with two Landau bands in the basis to illustrate the evaluation of the Hofstadter spectrum going far above the approximation of the Harper's Hamiltonian \cite{CastroNeto}. As is known the Harper Hamiltonian describes the motion of electron in a discrete 2D lattice in the transverse homogeneous magnetic field in the frame of the approach of hopping parameters. In other words magnetic field does not effect on the quantization strength in each QD and on the tunneling between the QDs but only on the phase shifts of the wave function due to the translations from one cite of SL to another. It means that the results obtained in our work would approach to ones obtained in the framework of Harper Hamiltonian for small enough values of magnetic field and when there is no mixing between the Landau bands due to the SL potential. For a honeycomb lattice this conditions can be fulfilled taking only two Landau bands in the expansion of the wave function as a minimal basis for the description of two ``graphene-like" minibands. Indeed, the right half (where the magnetic field is comparatively small) of the energy spectrum in the upper panel of Fig.\ \ref{Hof_nL1} is very similar with the known Hofstadter spectrum of graphene. However, with the increase of magnetic flux (with decrease of $\Phi_{0}/\Phi$) the energies undergo an up-shift due to the quantizing effect of magnetic field. Comparing the upper and the lower panels in Fig.\ \ref{Hof_nL1} or in Fig.\ \ref{Hof_nL5} one can observe an opening of a gap in the graphene-like Hofstadter spectrum due to the ellipticity of QDs and with an oscillating width along with the change in the magnetic flux. As is seen from Fig.\ \ref{Hof_nL5}, the Hofstadter-like spectrum is ``deformed" over all the considered range of the values of magnetic flux. One can observe here larger energy gaps for smaller values of magnetic flux when one considers six Landau bands in the expansion of the wave function per UC comparing with ones in Fig. \ref{Hof_nL1}. The slight up-shift of energies corresponding to SL with elliptical QDs (lower panels in Figs.\ \ref{Hof_nL1} and \ref{Hof_nL5}) is in accordance with the results shown in Fig.\ \ref{DispEllipt}.

In Fig.\ \ref{Magn} the magnetization in honeycomb SLs which are composed of circular (upper panel) and elliptical (lower panel) QDs versus inverse magnetic flux (in the unites of inverse flux quantum) is presented. The calculations are performed for 18 different rational values of $\Phi_{0}/\Phi$. These values are mentioned by vertical dashed lines, while the values of $\Phi/\Phi_{0}$ are mentioned near the graphs. We consider low temperatures ($1$K, $2$K and $3$K), so only the 1st and the 2nd minibands have significant contribution in the magnetization. As is obvious from the figures, the magnetization is always negative, that is the system is a diamagnetic. Generally, magnetization undergoes strong oscillations which are especially pronounced in the mid values of the magnetic flux. These oscillations are connected with the $pq$-fold splitting of the Landau bands in subbands with smaller widths and with the change of the periodicity of the system depending on $h_{1}$ and $h_{2}$. Interestingly, the magnitude of the magnetization is comparatively larger for fractional values of the flux compared to its integer values. This is a consequence of almost flat minibands with very weak dispersion at fractional valiues of $\Phi/\Phi_{0}$. Moreover, for even values of $\Phi/\Phi_{0}$ the magnitude of the magnetization is less than for its odd values. This is because of the degenerated Landau orbitals in a UC of SL mutually compensate each other. When $pq=4$, the four Landau orbitals in a UC are almost totally compensated in SL with circular QDs and the magnetization is nearly zero (see the upper panel of Fig.\ \ref{Magn}). The elliptical shape of QDs makes the effect of compensation weaker leading to non-zero magnetization for the same value of $\Phi/\Phi_{0}$ (see the lower panel of the figure). For large enough values of magnetic flux the values of the magnetization magnitude decrease and its oscillations weaken. This is due to the vanishing values of the thermal distribution function corresponding to the higher values of electron energy. Note, that the effect of temperature on the magnetization significantly depends on the magnetic flux per UC. Namely, for the SL with circular QDs the increase of the temperature leads to an obvious increase of the magnitude of magnetization for $\Phi/\Phi_{0}=$ $7/6$, $5/4$, $4/3$, $7/2$ and to its decrease for $\Phi/\Phi_{0}=$ $6/5$, $5/3$, $7/4$, $7/3$, $3$, $5$, $6$, $7$. One can also note that the arrangement of the values of magnetization corresponding to different values of temperature in a SL with elliptical QDs differs from ones in a SL with circular QDs when $\Phi/\Phi_{0}=6/5$ and $\Phi/\Phi_{0}=7/5$.      

\section{Conclusions}
\label{conclusions}
\vspace{-0.1cm}
Summarizing, we present a comparative study on electron energy dispersions and the magnetization of artificial graphene-like honeycomb SL composed of cylindrical and elliptical QDs.
We develop our theoretical model in the frame of the method proposed earlier by Ferrari, where a complete orthonormal set of basis wave functions is used, which reflects both the SL translational symmetry and the wave function phase-shifts due to the transverse magnetic field in the symmetric gauge of the vector potential. 
Our calculations indicate a topological change in the miniband structure due to the ellipticity of QDs.
We observe non-trivial displacements in the reciprocal space of the energy dispersion surfaces and transformations in the translational symmetry of the system when passing through different rational values of the number of magnetic flux quanta per UC of the SL.
The maxima of the dependencies of the DOS are duplicated due to the ellipticity of QDs for some values of the magnetic flux.
The Hofstadter spectrum of the SL with circular QDs qualitatively coincide with one for graphene for comparatively small values of magnetic flux and when two Landau bands are considered in the expansion of the wave function. However, the consideration of higher Landau bands leads to a significant modification of the Hofstadter spectrum. The ellipticity of QDs leads to a gap opening and to a considerable modification in the Hofstadter spectrum.
The magnetization reveals non-trivial oscillations depending on the magnetic flux. The fact of the magnetic flux being integer or fractional plays a crucial role in the diamagnetic behaviour of the system. The oscillations in the magnetization, as well as the arrangement of its values corresponding to different values of temperature considerably depend on the geometry of QDs.

\section{Acknowledgement}
This work was financially supported by the Armenian
State Committee of Science (grants No 21SCG-1C012,
No 21T-1C247, 20TTWS-1C014 and No 21AG-1C048), by the Research Fund of the University of Iceland, and the Icelandic Infrastructure Fund.

The computations were performed in the Center of Modelling and Simulations of Nanostructures at Yerevan State University.

\end{document}